\documentclass[twocolumn,showpacs,preprintnumbers,amsmath,amssymb,amsmath]{revtex4}

\usepackage{graphicx}
\usepackage{dcolumn}
\usepackage{bm}
\usepackage{hyperref}
\usepackage{epsf}
\usepackage{float} 
\newcommand{\eq}[1]{(\ref{#1})}
\newcommand{\la}{\label}
\newcommand{\bea}{\begin{eqnarray}}
\newcommand{\eea}{\end{eqnarray}}
\newcommand{\beq}{\begin{equation}}
\newcommand{\eeq}{\end{equation}}
\newcommand{\be}{\begin{equation}}
\newcommand{\ee}{\end{equation}}
\newcommand{\ii}{{\rm{i}}}

\newcommand{\vv}{{\rm v}}
\newcommand{\uu}{{\rm u}}

\newcommand{\p}{\partial}

\def\XXint#1#2#3{{\setbox0=\hbox{$#1{#2#3}{\int}$ }
\vcenter{\hbox{$#2#3$ }}\kern-.5\wd0}}

\usepackage{amsmath}
\usepackage{amsfonts}
\usepackage{varioref}

\begin{document}

\title{Quantum Hydrodynamics of Fractional Hall Effect:\\ Quantum Kirchhoff Equations}

\author{P. Wiegmann }
 \affiliation{ Department of Physics, University of Chicago, 929 57th St, Chicago, IL 60637}
\date{\today}
%

\date{\today}

\begin{abstract}
We argue that  flows of the quantum electronic liquid in the Fractional Quantum Hall state are comprehensively described by the hydrodynamics of vortices in  the quantum   incompressible rotating liquid. We obtain the quantum hydrodynamics of vortex flow by quantizing   Kirchhoff equations for vortex dynamics.  We  demonstrate  that quantized Kirchhoff equations capture all major features  of FQH states including  subtle  effects  of  Lorentz shear force, magneto-roton spectrum, Hall current in a non-uniform electromagnetic field, thus providing a powerful framework to study FQHE and superfluids.\end{abstract}

\maketitle

\noindent\emph{1. Introduction.} In  Fractional Quantum Hall (FQH) regime  electrons form a peculiar quantum liquid.  Characteristic features  of the liquid are:  flows are incompressible \cite{L},   almost  dissipation-free \cite{T,T1},  the Hall conductance  is  quantized 
\cite{T1},  vortices are elementary excitations, vortices carry  fractionally quantized negative electronic charges \cite{L}, they are separated from the ground state by a gap \cite{T,GMP}. More subtle features are   the universal  anharmonic term of the structure factor and magneto-roton minimum in excitation spectrum \cite{GMP} (see \eq{s}), quantized   double layers of a density and shear  at  boundaries and vortices \cite{W}, the Lorentz shear force \cite{HV}. 

Such liquid does not posses  linear gapless waves except edge modes propagating along the boundary. Only available bulk flow is a non-linear flow of vorticity. Since in  FQH regime vorticity is linked to the electronic charge, flows of vorticity are   related to a charge flow \cite{electrostatics}. 

A natural approach to study flows of FQH states is hydrodynamics.  Not all quantum liquids are subject of quantum hydrodynamics. Quantum hydrodynamics is based on a fundamentally restrictive assumption    that all quantum states  are fully characterized by the density and velocity. Apart from  of superfluids (and  superconductors) and Luttinger liquids,  FQH liquid is, yet, another
important case.

Hydrodynamics of quantum
fluids  goes back to Landau \cite{Landau} and
Feynman \cite{Feynman}, but remains in its infancy.  A quest for the hydrodynamics of FQH liquid has been   originated by a seminal paper \cite{GMP}, further discussed in \cite{Kivelson,Dung,Read89,Stone}, it  is  a focus of  a renewed  interest. 

 As always in hydrodynamics only  a few basic principals, symmetries,  and a few phenomenological parameters are sufficient to formulate   fundamental equations. To this aim an underlying  microscopic Hamiltonian describing emergent  FQH states  under a strong Coulomb interaction is in fact not necessary. 

In the letter we attempt to formulate these  principals and to develop  hydrodynamics
of \! FQH bulk states in a  close analog of Feynman theory of superfluid helium\! \cite{Feynman} and magneto-roton theory of collective excitations  by \cite{GMP}.

We argue that flows of FQH liquid are equivalent to  flows of vortices in the quantum incompressible rotating Euler liquid. Based on this  conjecture we obtain
the major  features od FQHE including subtle effects of Lorentz shear force \cite{HV}and magneto-roton spectrum \cite{GMP} missed by the previous hydrodynamics approaches  \cite{Kivelson,Dung,Read89,Stone}.

Hydrodynamics of   vortex flows (classical and quantum alike) is an interesting subject by its own. Apart from FQHE it is also relevant to the theory of superfluids and classical hydrodynamics. 

Quantization of hydrodynamics is a subtle matter. It is best achieved through quantization of
Kirchhoff equations \cite{C}.
Classical  Kirchhoff equations describe a motion of vortices in a 2D  incompressible isentropic fluid.  We show that, quite remarkable,  quantum Kirchhoff equations  capture all  known features of  FQH liquid. They can be used as a platform for studying  FQH Effect.

In this paper we 
consider only the simplest Laughlin's cases where  fraction \(\nu^{}\) is an inverse of an odd integer, say 1/3. In these cases  electronic liquids do not posses  additional symmetries. Extension to other FQH states, not discussed here,  is  possible and  interesting.

We start by reminding classical Kirchhoff equations for rotating incompressible Euler flows  (see e.g.,\cite{Newton}).
\smallskip

\noindent\emph{2. Classical Kirchhoff equations.} 2D incompressible isentropic  flow (that is a flow  where gradient of density  are
orthogonal to gradients of pressure, or zero) is fully characterized by its vorticity. Vorticity obeys a single (pseudo) scalar equation, which in the case of inviscid  fluid has a simple geometrical meaning: the material derivative of the
vorticity vanishes  
\begin{align}\la{E}{D_t\varpi}\equiv\left(\frac{\partial}{\p t} +
  u\cdot \nabla\right)\varpi=0.
\end{align}In other words vorticity \(\varpi= { \nabla}\times
  u\) is  transported
 along    divergent-free velocity field \(  u\) \begin{align}\label{i}
 \nabla\cdot{  u}=0.
\end{align} Helmholtz, and  later Kirchhoff realized that there is a class of solutions of the vorticity equation \eq{E} which consists of a finite number of  point-like vortices moving  by Magnus forces. In this case  the complex velocity \(\uu=u_x-iu_y\)  is a meromorphic function 
 \begin{align}\label{k}
\uu(z,t)=-\ii \Omega\bar z+\ii\sum^N_{i=1 }\frac{\Gamma_i}{z-z_i(t)} ,
\end{align}
where  \(\Omega\) is an angular velocity  if the fluid is rotated, \(N\)   is a   number of vortices,  \(\Gamma_i\) and   \(z_i(t) \) are circulations and positions
of vortices. We  denote
 \(z\!=\!x\!+\!\ii y,\p\!=\!\frac{1}{2}(\nabla_x\!-\!\ii \nabla_y) \) and use
the roman script for complex vectors \(\mathrm{a}=a_x-ia_y\).  

A substitution this "pole Ansatz" into the Euler equation \eq{E}
yields that the number of vortices \(N\) and the circulations    
\(\Gamma_i\) do not change in time,   but moving position
of vortices \(z_i(t)\) obey the  Kirchhoff equations.
 
If the rotation of fluid is very strong, vortices  prefer to be of 
the   same sign opposite to the direction of the rotation. Bearing in mind the quantum case we assume that vortices have the same (minimal) circulation  \(\Gamma_i=\Gamma\).  
Then Kirchhoff equations obtained by
localizing  the Euler equation to   vortex cores at  \(z=z_i\) expresses
velocities \(\vv_i\) of  vortex cores through their positions \begin{align}\label{u}
\vv_i\equiv  \dot {\overline z}_i=-\ii \Omega\bar z_i+\ii\sum^N_{i\neq j }\frac{\Gamma}{z_i(t)-z_j(t)}.
\end{align}
The Kirchhoff equations replace the non-linear PDE \eqref{E} by the  dynamical  system, reflecting integrability of the Euler flow. They can be used for different aims. Equations describe chaotic motions of a finite number of vortices if \(N>3\). Alternatively they can be used  to  approximate virtually any flow if \(N\to\infty\). 
\smallskip
 
\noindent\emph{3.  Chiral flow.} The flow  relevant for FQHE is the {\it chiral flow,} where a large number of background vortices    uniformly distributed with the mean density  \(\bar\rho=\/\Omega/(\pi \Gamma) \)  largely compensates rotation.  This  is a very special flow in fluid mechanics. There we distinguish two types of motion: fast motion of the fluid around vortex cores, and a slow motion of vortices.  In this respect vortices  themselves must be considered as a (secondary) fluid. At the ground state of chiral flow  vortices do not move, but the fluid does. Vortices  carry an inertia \(m_*\) -- a phenomenological parameter  determined by the equation of state. It is not directly related to the inertia of fluid.

  In this paper we propose to model FQHE by a quantized slow flow of vortex fluid. \smallskip
 \smallskip

\noindent\emph{4. Quantization of  Kirchhoff equations.}
 Quantization  of the Kirchhoff equations consists of three steps:  canonical quantization,
a choice of the representation and  the inner product. 

It is convenient to use a circulation \( q=m_*\Gamma \) of  momenta of  vortices  \(m_*v_i\). The Poisson brackets followed from canonical Hamiltonian structure of the hydrodynamics. They are equal to  the volume per particle per circulation \(\{x_i,\,y_j\}_{P.B.}=1/(2\pi\bar\rho q)\). We replace them  by commutators \begin{align}\label{C}
\{\bar z_i,\,z_j\}_{P.B.}\rightarrow[\bar z_i,\,z_j]=2\ell^2\delta_{ij}.
\end{align}
Here \(2\ell^{2}=\nu/(\pi\bar\rho )\) has a dimension of area. Dimensionless number
\(\nu^{}=\hbar/q\)\(\) is a semiclassical parameter.

The next step is a choice of states. We assume that states are  holomorphic polynomials of \(z_i\). Then operators
\(\bar z_i\)  are  canonical momenta \be\la{r}\bar  z_i  =2\ell^2\p_{z_i}.\ee The third step is to specify the inner product. We impose  the \emph{chiral condition}: operators \(\bar z_i\)  and \(z_i\) are assumed to be Hermitian conjugated
\begin{align}\la{hermitte}
\text{\small chiral condition}:\quad &\bar z_i=z_i^\dag.&
\end{align} This condition combined  with representation \eq{r} identifies a set of  states with the Bargmann space \cite{Bargmann}:\ the Hilbert space  of analytic polynomials \(\psi(z_1,\dots,z_N)   \) with the inner product (\(\exp(-{z\bar z})\bar \p \exp{(z\bar z)}=z\) insures
(\ref{r},\ref{hermitte}))\begin{align}\la{B}
\langle\psi'|\psi\rangle=\int d\mu
\overline{\psi^\prime}\psi,\quad  d\mu= \prod_ie^{-\frac{|z_i|^2}{2\ell^2}}{d^2z_i.}\end{align}
\!\!
Consequently \(\bar z_i^\dag=-2\ell^2\bar\p_{z_i}+z_i\). The normal ordering in the Bargmann space assumes  that holomorphic operators stays to the left  to anti-holomorphic operators.

Eqs. (\ref{u},\ref{r}) help to write velocity operators of vortices\begin{align}\la{v}
   \Gamma^{-1}\vv_i=-\ii2\nu{ \p}_{z_i}+\ii\sum_{i\neq
j}\frac{1}{z_i-z_j}, \quad 
\end{align}
 Eqs.(\ref{u}-\ref{v}) are quantum chiral Kirchhoff equations.
They are readily  generalized  to a sphere,  or a torus.
\smallskip

\noindent\emph{5. Quantum Chiral Kirchhoff Equations and FQHE. }We  identify quantum  chiral Kirchhoff equations with FQHE. 

First we comment  that Bargmann space of analytic polynomials is  another way to say  that all states belong to the lowest Landau level \cite{Bargmann} (and also \cite{GMP,GJ}).
Then we assume that a state where  vortices are at rest is the  ground state of the system $\psi_0$. It
nulls  all velocity operators  \( \vv_i\,{\rm Arg}\,[\psi_0]=0\) (velocity operator acts on the phase of
the w.f.).
A common solution of  the  set of  1st order PDEs  is the holomorphic part of the Laughlin w.f.  \begin{align}\la{L} \psi_0=\prod_{i>j}(z_i-z_j)^\beta,\quad\quad\hbar \beta=q.\end{align}   
The w.f. becomes single valued if \(\beta\) is integer, antisymmetric if \(\beta\)  is an  odd-integer.

In this interpretations vortices are identified with "particles"  entered
into the Laughlin function. Hence  a  quasi-hole \(\psi_h= \prod_i(z_0-z_i)\psi_0\)
\cite{L} is a hole in the uniform  background of vortices - an anti-vortex.
 Thus we  assign  the electronic charge to vortices and identify angular
velocity with the cyclotron frequency of vortices \(\Omega\!=\!eB\!/\!(m_*c)\).
Then the
 entries of the Kirchhoff equations are  the  magnetic length  \(\ell\!=\!\sqrt{\!\hbar
c/\!e B}\) and  the filling fraction
\(\nu\!\!=\!\bar\rho hc/e\!B\!=\!\hbar/q\) (\(\Gamma=\Omega/(\pi\bar\rho)\).

The  phenomenological parameter \(m_*\) is the inertia 
   of the vortex.   It  is  naturally to assume that the energy associated
with the inertia
 \(\Delta_\nu=\hbar\Omega=\hbar^2/(\ell^2 m_*)\) is of  the same order  as the   energy of a quasi-hole at rest,  or equivalently, is a gap in the excitation spectrum.  The latter is of the order of Coulomb energy \(e^2/\ell\). It  is  known experimentally  \(\Delta_\nu\sim 10K\) \cite{T}.  The very existence of the FQH state requires that this energy scale must be less than the cyclotron frequency \(\hbar\omega_c\sim 25m eV\gg\hbar\Omega\), or that \(m_* \) 
 exceeds the bare electronic mass \(m_*\gg m_e\). An assumption that the energy of the quasi-hole at rest is the  gap and is related to  the  inertia of the flow is the major physical input led to the hydrodynamics of FQHE.

A noticeable  feature of the identification is that slowly varying  external fields (potential well,  gradients of temperature etc.) are coupled to slow moving vortices. For example a potential well \(U(r) \) adds to the energy  \(\sum_i U(r_i)   \), where \(r_i\) are coordinates of vortices, not fluid particles. It yields the Lorentz force  \(-i[U,\bar z_i]= i2\ell^2\p_{z_i} U \) into  the r.h.s. of \eq{v}. In the flow which keeps the center mass at the origin, Kirchhoff equations yield  \(\sum_i\vv_i=\ii (\pi q\bar\rho)^{-1}\sum_i\p_{z_i} U\). This  implies   the fractionally quantized Hall conductance 
\(\sigma_{xy}^{}=e^2/(2\pi q)=\nu (e^2/h)\).

Finally we emphasize that the   velocity \eq{v} is different than velocity of individual
electrons and coordinates of their guiding centers.

In the rest of the paper we show that Kirchhoff equations contain other, more subtle properties of FQHE. To this end we must develop   the hydrodynamics  of quantum vortex flow. To the best of our knowledge this has not been done even for the classical fluids. \smallskip
 
\noindent\emph{6. Velocity field of the flow of vortices.}  Eulerian hydrodynamics
of vortex computes the flux of vortices
\begin{align}\la{p}
  P=\!\frac{m_*}{2}\!\sum_{i}\!\{\delta(r\!-\!r_i),  v_
i\}\!=\!\frac{m_*}{2}\{\rho,   v\}\!=\!m_*\sqrt\rho\,   v\, \sqrt\rho, \end{align}
where
\(r_i,  v_i\)  are coordinates and  velocities  of vortices,   \(\{,\}\)
is the anti-commutator and  \(\rho(r)=\sum_{i}\delta(r-r_i)\) is the  density
of vortices.
This formula also defines the velocity of the vortex flow \(  v(r)\).
 
By construction operators \(\mathrm{P   }\) and \(\mathrm{P^\dag}\)
are ladder operators. They annihilate the ground  state \(\mathrm{P\psi_0=\bar\psi_0\mathrm{P}^\dag=0}\).

 In  hydrodynamics the mass density
 of the fluid and  mass flux of the flow are   independent fields obeying   the   continuity and  the Euler equations.
  Contrary, the velocity of the vortex flow \(  v(r)\)   (as well as the velocity of the fluid \(  u(r) \)) are expressed through the  density of vortices
 \(\rho(r).\)
   We must determine these relations. Before we proceed the comment is in order.

 In 2D
Euler liquid the  mass density
of the fluid and mass density of vortices, both, obey the continuity equations. Hence
an initially  imposed local condition, \(n(r)=m_*\rho(r)\)
is compatible with the dynamics at all   time.    This condition 
reduces the set
of states being essentially equivalent
to the chiral constraint \eq{hermitte}.  Thus the vortex
flow can be formally considered as   a special flow
of 2D Euler liquid with a constituency relation between
the  flux    and the  density of the liquid 
  \cite{comment}. From this point of view  \(  J=\frac{1}{2}\{n,  u\}=\frac{m_*}{2}\{\rho,   u\}\) is the
mass flux of the special flow. It includes fast fluid motion around  vortices and
a slow motion of vortices.  This is  a useful but  an auxiliary object. Instead we are after the flux of the vortex flow \eq{p}, which consists of only slow motion of vortices.

The flux of the vortex flow  subtly differs to the flux of the fluid.
We compute them now.   

The velocity  of the fluid is  easy obtain. It is given by  \eq{k}, where we represent the operator \(\bar z\) by \eq{r}, use   \(\rho\nabla 
\frac{\delta}{\delta\rho}=\sum_i\delta(r-r_i)\nabla_{r_i}\), and pass to the continuum limit 
  \begin{align}\la{U}
m_*\uu=2  \p\pi_\rho+\ii q \int\frac{\rho d^2\xi }{z-\xi},\quad \pi_\rho\!=\!-\!\ii\hbar 
\frac{\delta}{\delta\rho}.
\end{align}  
The flux \(J=\frac{m_*}{2}\{\rho,
  u\}\) follows. Computing the flux of the vortex flow \eq{p} we use: the \(\bar\p\)-formula
\(\pi\delta=\bar\p(\frac 1z) \)  and the identity  \(2\sum_{i\neq
j}\frac{1}{z-z_i}\frac{1}{z_i-z_j}=(\sum_i\frac{1}{z-z_i})^2-\sum_i(\frac{1}{z-z_i})^2\).
With the help of \eq{v} a simple computation yields \begin{align}&\!\mathrm{P}=\!\{ \rho,\p
\pi_\rho\}\!+\!i\frac{q} {2\pi}\bar\p[(\sum_i\frac{1}{z-z_i})^2\!-\sum_i\!\frac{1}{(z-z_i)^2}]\nonumber
\\&=\{ \rho,\p
\pi_\rho\}+{\ii \rho}\sum_j\frac{q}{z-z_j}+\ii\frac q 2\p\rho=\mathrm{J}+\ii\frac
q 2 \p\rho\label{v1}. 
\end{align}
We obtain the important relation between flux of the vortex flow and the flux of the fluid. Using the notation \( a^\ast_\mu=\epsilon_{\mu\nu}a_\nu\) for 2-vectors we obtain
 \begin{align}\la{shift}
  P=  J+\frac {q}{4}  \nabla^\ast\rho,\quad    v=  u+\frac {\Gamma}{4}\rho^{-1} \nabla^\ast\rho,
\end{align}
 The shift \eq{shift} holds in the classical and  the quantum cases.  
It  has
 far reaching consequences \cite{Calogerocomment}.

The shift \eq{shift}  can be seen as  a similarity transformation. It preserves
the volume.  Hence the flow of vortices is  incompressible
 \(  \nabla\cdot   v=0 \) like the fluid itself.

The shift has a simple meaning. Velocity of  the fluid \(u\) diverges at a core of an isolated vortex
 (as is in (\ref{k})). However, velocities of  vortices are finite.
The  shift  removes that singularity.

Further  meaning of the shift   is  seen from monodromy of the wave function.
Monodromy is the circulation  of each particle  in units of \(\Gamma\). It
is equal   to the number of magnetic flux quantum
piercing the system \(N_\phi\). The circulation of  a particle (a vortex)
around the system of remaining $N-1$ vortices is  \( \Gamma(N-1)=2\pi\oint
  v\cdot d  r \). The monodromy is \(N_\phi=\beta(N-1)\), i.e., the number
 of zeros of the w.f. with respect
to each coordinate. On the other hand the circulation  \(2\pi\oint   u\cdot
d  r\)
gives  the total charge \(\Gamma N\). The  shift  amounts for the difference.
It  simply means that a vortex does  not interfere with itself. 
 Eq. \eq{shift} can be seen as a local version of the global condition
\( N_\phi= \beta N-2\bar s\), where the shift \(\hbar\bar s=\!\!q/2\) \cite{shift2}.

We summarize the formulas for flux and velocity \begin{align}\la{p1}
&  P=\frac 12\{ \rho,  \nabla\pi_\rho\}-\frac{q}{2}\rho  \nabla^\ast\varphi+\frac{q}{4}  \nabla^*\rho,\\
& m_*  v=  \nabla\pi_\rho-\frac{q}{2}  \nabla^\ast\varphi+\frac{q}{4} 
\nabla^*\log\rho.\la{vv}
\end{align}
Potential
\(\varphi\)  
obeys the Poisson equation  \(\Delta
\varphi\!=\!-\!4\pi(\!\rho\!-\!\bar\rho)\). It is chosen  such that    flux
vanishes
at the ground state.

\smallskip

\noindent\emph{7. Chiral constituency relation.}
The next step is to express the flux of the vortices  in terms of  their density. 

Using the formula \([\rho,\p
\pi_\rho]=
 -\ii\hbar\p\rho\) and the chiral relation ${2\ell^2\p}_{\!z_i}^\dag=-2\ell^2\overleftarrow{\p}_{\!z_i}+\bar z_i$, we write\begin{align*}
\{ \rho,\p
\pi_\rho\}=-2\p
\pi_\rho^\mathrm{T}\rho+[ \rho,\p
\pi_\rho]=-\ii\hbar(
\p+\frac{\bar
z}{2\ell^2}\!)\rho.\end{align*}
Applying this formula to (\ref{p1}) we obtain the chiral constituency
relation. We write it in two suggestive forms
\begin{align}
&   P =-\rho  \nabla^\ast\Psi,\quad\Psi= \frac{q}{2}[\varphi-(\frac{1}{2}-
 \nu)\log\rho],\la{v3}\\
&\mathrm{P}=\frac{\ii q}{\pi}\bar\p\mathcal{T},\quad \mathcal{T}=\frac{1}{2}(\p\varphi)^2-(\frac12-\nu)\p^2\varphi.\la{v4}
\end{align}     
 The field \(\Psi\) has a meaning of the stream function, hence vorticity is \begin{align}\omega\!=\!-\Delta\Psi\!\!=\!2\pi q[\rho-\bar\rho+\frac{ 1}{4\pi}(\frac{1}{2}- \nu)\Delta\log\rho].\end{align}
 We
observe that vorticity  of the chiral flow   differs from the density
of vortices by the term  \(\propto \Delta\log\rho\). The coefficient
comprises of the shift and quantum correction.
The source of quantum corrections is the difference of the vorticity of the fluid \(\varpi=2\pi q[\rho-\bar\rho-\frac{
\nu}{4\pi}\Delta\log\rho]\)
from the density of vortices. 

 Integration of   \eq{v3} gives the global version of the chiral condition.
It connects the
 moment of inertia  \(\mathrm{L}=\int(  r\!\times\! 
 P)d^2r\) and the
 gyration 
\(\mathrm{G}=\int\! r^2(\rho-\bar\rho)d^2r\) of the flow: 
\begin{align*}\ell^2\mathrm{L}=\hbar
\mathrm{G}+N\ell^2\hbar(\beta-2)\end{align*}
This form  generalizes the   exact sum rule of  the Laughlin
state
at  \(P=0:\quad
\sum_i \langle 0||z_i|^2|0\rangle=N\ell
^2\left(N-(\beta-2)
\right).
\)

Eq. (\ref{v3})  can be used to find  density  profiles for various coherent states \cite{correltations}. Consider e.g.,   a quasi-hole \(\psi_h= \prod_i(z_0-z_i)\psi_0\) \cite{L}. This state describes an anti-vortex (a hole
in the sea of  vortices). Its flux is \(\mathrm{P}=\ii\hbar \frac{\rho}{z-z_0}\). Eq \eq{v3}  expresses its density through the 2-points function.
Eq.\eq{s1}  computes the change of the gyration by a quasi-hole: \(N^{-1}\int r^2\delta\rho  d^2r=-\ell^2=-\frac{\nu}{2\pi\bar\rho}\). This result often interpreted as a fractional charge of the quasi-hole - it occupies a fraction of volume per particle. Outside of the  core,   the quasi-hole as a source for vorticity in a classical equation \cite{comment1}
\begin{align*}-\delta(r-r_0)= \beta[\rho-\bar\rho+\frac{ 1}{4\pi}(\frac{1}{2}-
\nu)\Delta\log\rho]. \end{align*}

%
\smallskip

\noindent\emph{8. Governing equation for the vortex flow.}
  Since the density of vortices determines its flux, the continuity equation \(m_*\dot\rho+  \nabla\cdot   P=0\) is the only governing equation of the chiral flow. Eq.(\ref{v3})
provides the close form 
\begin{align*}\dot\rho+\frac \Gamma 2\nabla  \varphi\times  \nabla\rho=0.\end{align*} This is of course the standard equation for vorticity of  incompressible liquid. Peculiar  effects are burred  on  boundaries or in initial data. A moving edge must be a stream line  \eq{v3}. Since stream lines are non-linearly connected to level lines of the potential \(\varphi\)   
the dynamics  differs  from  that of a free boundary  of  incompressible fluid. In \cite{W} we showed that edge dynamics is governed by Benjamin-Ono equation. Similarly, initially given vorticity is non-linearly related to the density of vortices, hence its dynamics differs  from Euler flows.

\smallskip

\noindent\emph{9. Hamiltonian structure of  vortex dynamics}.
Kirchhoff
equations \eq{u} are integrable. Same equations   follow from  different Hamiltonians \cite{KirchhoffHamiltonian}.
One  Hamiltonian comprises by 
 kinetic energies of the fast fluid
 motion and  slow drift of vortices 
 \(
H_E=\frac {1}{2m_*}\int \mathrm{J}^\dag \rho^{-1}\mathrm{J }
d^2r\).    
An equivalent form of the  classical version of this  Hamiltonian (but without incompressible and chiral conditions), has been  proposed
    as a description of FQHE in \cite{Kivelson, Read89,Dung, Stone}. Our main result is that flows of FQH states are described by another Hamiltonian (used 
in  \cite{W}, see also \cite{Feinberg}),
containing only slow motion of vortices \cite{normalordering,Jack} \begin{align}{H}=\frac
{1}{2m_*}\int\mathrm{P}^\dag\rho^{-1}\mathrm{P}\ d^2r.\la{V}
 \end{align}
  Both yield the same Kirchhoff equations but have different energy. The difference   is essential. The latter yields correct properties of the Laughlin state reflected by \eq{shift}, the magneto-roton minimum  in the excitation spectrum  of \cite{GMP} and  Lorentz shear force discussed below. The former does not \cite{commentK,commentCS}.

With the help of the chiral relation \eq{v3} we obtain 
 \begin{align}
 {H}=\frac{1}{2m_*}\int\rho|  \nabla\Psi|^2 d^2 r,
 \end{align} 
 where   the stream function is given by 
\eq{v3}. 

The symplectic structure
can be expressed in two forms: in terms of  velocity or flux 
of the fluid,  or in terms  of the vortex fluid. The first structure  is the  canonical Heisenberg current algebra  
followed from  \eq{U} \begin{align}\la{JJ}
{m_*^2}[u(r),\,u^\dag(r')]=4\pi q \delta(r-r').
\end{align}The second   is an extended  algebra obtained by the
transformation \eq{shift} \cite{bi-hamiltonian,Virasoro}. It is conveniently written in terms of the flux
\begin{align}
&[\mathrm{P}(r),\!\mathrm{P}^\dag(r')]\!=2\hbar[-  P\!\times\!  \nabla\!+\!
q\rho(2\pi \rho\!+\!\frac 14\Delta)]\delta(r\!\!-\!\!r').\la{PP}
\end{align}
The advantage of the first structure
is its canonical form. The advantage of the second structure is  that operators
\(\mathrm{P}\) , \(\mathrm{P^\dag}\) are ladder operators annihilating the ground state.

The Hamiltonian \eq{V} and the algebra \eq{PP} acting in the Bargmann space are the compact and comprehensive formulation of FQHE.\smallskip

\noindent\emph{10. Structure function and excitation spectrum.}  We illustrate the current algebra by computing the celebrated results of \cite{GMP} for the structure factor and the excitation spectrum. Let   \(\rho_k
\) is a Fourier mode of a small density  modulation. In the linear approximation  \eq{v3} connects it to the flux  mode   \(\mathrm{P}_{
k\!}=\!\hbar \frac{2{\rm k}}{(k\ell)^2}(1\!-\!\frac{\beta-2}{4}(k\ell)^2)\rho_k\). Then project  \eq{PP} onto the ground
state and compute its  r.h.s.. This gives  the correlation of  flux modes \(\langle \mathrm{P}_k\mathrm{P}^\dag_{k}\rangle=2\bar\rho\hbar^2\ell^{-2}(1\!-\!\frac{\beta-2}{4}(k\ell)^2)  \).  Comparing we obtain the Feynman-Bijl formula \cite{Feynman} for  the single mode approximation of the excitation spectrum \(\Delta(k)\!\equiv\!\frac{\langle
\mathrm{P}_k\mathrm{P}^\dag_{k}\rangle}{2m_*\bar\rho^2}\), and  the structure factor  \(s(k)=\langle\rho_k\rho_{-k}\rangle\)\begin{align}\la{s}
\Delta(k)\!\!=\!\frac{\hbar^2 k^{2}}{2m_*s(k)},\quad s(k)\!=\!\frac{\bar\rho}{2}\frac{(k\ell)^2}{1\!-\!\frac{\beta-2}{4}(k\ell)^2}.
\end{align} 
  The excitation energy
\(\Delta(k)=2\pi\beta\frac{\hbar^2}{m_*}(1\!-\!\frac{\beta-2}{4}(k\ell)^2)\) shows the magneto-roton minimum of  \cite{GMP}, the leading order of the structure factor   also agrees with  
 \cite{GMP}.
\smallskip

\noindent\emph{ 11. Stress tensor and Lorentz shear force.}
The shift \eq{shift} causes  subtle hydrodynamic phenomena. One of them is the Lorentz shear force,  a conservative force acting normal to the shear flow. 
It enters to the conservative part of the viscous tensor, which we now compute. Initially being introduced for the integer QHE in  \cite{Avron} and extended to FQHE in   \cite{Tokatly,ReadHV}, the Lorentz shear force is in fact, the classical phenomena. It is a feature of vortex flow.
 
  We  cast hydrodynamics  equations in the form of the conservation law
with the Lorentz force  \(  F\!=\!e  E\!-\!\frac
ec B  v\,^\ast\) \begin{align}\la{T}
\p_t P_\mu+\nabla_\nu \Pi_{\mu\nu}=\rho F_\mu,
\end{align}
 and    compute the  momentum flux \(\Pi_{\mu\nu}\). 

 We start from the first Hamiltonian structure  writing conservation law for the fluid flux \(J\):
\(\p_t J_\mu+\nabla_\nu \tilde \Pi_{\mu\nu}\)=0. 
In this case the  canonical property of the first structure  \eq{JJ} determines the form of the momentum flux:    
 \(\tilde \Pi_{\mu\nu}=  J_\mu(m_*\rho)^{-1}J_\nu\!+\!\rho\tilde p\delta_{\mu\nu}\), while  incompressibility condition  determines the  intrinsic pressure  \(\tilde p\) (all operator products are normally ordered). 

We obtain the  momentum flux of  vortices \(\Pi_{\mu\nu}\)  by the similarity transformation  \eq{shift}. Taking time derivative of \eq{shift} and using  the continuity equation, we write \(\dot
J_\mu\!-\!\dot
P_\mu\!=\!\frac {q}{4}\nabla^\ast_\mu(v\cdot\!\nabla)\rho\!=\!\!\nabla_\nu (\Pi_{\mu\nu}\!-\!\tilde
\Pi_{\mu\nu}).\) This gives the transformation of the momentum flux  \(\tilde\Pi_{\mu\nu}\!\to\!\Pi_{\mu\nu}=\\:P_\mu(m_*\rho)^{-\!1} P_\nu :-:\sigma_{\mu\nu}:\)  and the stress tensor \(\sigma_{\mu\nu}\)
\begin{align*}&\sigma_{\mu\nu}=-\rho p\delta_{\mu\nu}-\frac{q^2}{4}\nabla_\mu\sqrt\rho\nabla_\nu\sqrt\rho+\sigma'_{\mu\nu,}\\&\sigma'_{\mu\nu}=-\frac{ q }{4}\rho(\nabla_\nu v^\ast_\mu+\nabla_\nu^\ast v_\mu).\nonumber\end{align*}
The     viscous part of the   stress tensor, \(\sigma'_{11}\!=\!-\sigma'_{22}\!\!=-\frac{
q }{4}\rho(\nabla_xv_y\!+\!\nabla_yv_x),\;\sigma'_{12}\!=\!\sigma'_{21}\!=\!\frac{
q }{4}\rho(\nabla_xv_x\!-\!\nabla_yv_y)\) is the Lorentz shear force.  It is traceless, hence conservative. 

The  effect can  be interpreted in terms of semiclassical
motion of electrons.
A motion of  electrons consists of a fast motion along  small orbits  and a slow motion of  orbits. A shear 
flow strains orbits elongating them normal to the shear, boundaries and vortices.  Elongation
yields    an addition to the Lorentz, the Lorentz shear force which acts normal to the shear and proportional to the shear.  

An important consequence of the Lorentz shear force is accumulation of charges  on boundaries
and vortices -  {\it overshoots}. They govern dynamics of edge modes \cite{W}.\smallskip

\noindent\emph{ 12. Hall current.}
Another  consequence of the Lorentz shear force is the increase  of the Hall current by a non-uniform e.m. fields \cite{Son}. It follows from the conservation law \eq{T}  that  in the linear approximation the Lorentz force equilibrates by the  Lorentz shear force \( e  E-\!\frac
ec B   v\,^\ast\approx\frac q4\Delta   v\,^\ast\approx \frac q4c\Delta(  E/B)\). As the result the Hall current increases. In  a uniform magnetic
field 
\(  j_e=\sigma_{xy}\left(1+\frac{1}{4\nu}(k\ell)^2\right)  E_k^\ast\). 

So far we neglected diamagnetic effects by counting the energy from  the ground state. The ground state energy depends on the magnetic field as \(-\frac 12M\hbar\omega_c\bar\rho\), where \( M \) is an orbital moment
 per  particle. In a non-uniform density it yields the diamagnetic current \(  j_{\it dia}=-\frac{|e|\hbar}{m_e}M  \nabla^\ast
\rho \).
  The diamagnetic currents gives a similar contribution to the Hall conductance as
the Lorentz shear force. Indeed, the density of  the chiral flow is determined by the flux according to the
  Eq.\eq{v3}. In
the leading order    \( m_*\Delta   v\approx{2\pi q}
  \nabla^\ast\rho\approx m\Delta(c  E^*/B)\), hence \(  j_{\it dia}\approx-\sigma_{xy}\frac{m_*}{m_e}M\ell^2\Delta  E^\ast\).  Thus diamagnetic current merely changes  \(\frac{1}{4\nu}\to \frac{1}{4\nu}+\frac{m_*}{m_e}M\)
in the formula for the current. This result has been obtained in \cite{Son} where \(m_*\)was set to be equal to bare electronic mass \(m_e\).  A factor \(m_*/m_e\) offers an avenue to obtain the inertia \(m_*\) by measuring of   the increase of  the Hall current
by a non-uniform electric or magnetic  fields.  Comparing it with an independently measured gap allows to check Feynman-Bijl formula \eq{s}.    \smallskip

\noindent
%
%
 \smallskip
   
\noindent\emph{Acknowledgement.} This study had started as a common project with A. G. Abanov. It turned out that our methods and results were  complimentary,  that we decided to present them in separate publications, see \cite{Abanov}. 
Discussions of hydrodynamics of quantum liquids with I. Rushkin, E. Bettelheim and T. Can  and their help are acknowledged. 
 The author
 thanks the International Institute of Physics (Brazil) and Weizmann Institute of Science  for the hospitality during the
completion of the paper. The works was supported by NSF DMS-1206648, MRSEC  DMR-0820054 and BSF-2010345.
 

\begin{thebibliography}{10}
 \bibitem{L}R. B.Laughlin, Phys. Rev. Lett. 50, 1395 (1983).

\bibitem{T1}D. C. Tsui, H. L. Stormer, A. C. Gossard, Phys. Rev. Lett. 48, 1559 (1982).

\bibitem{T} R.R. Du, H.L. Stormer, D.C. Tsui, L.N. Pfeiffer, and K.W. West,
Phys. Rev. Lett. {70}, 2944 (1993).

\bibitem{GMP}S.
M. Girvin, A. H. MacDonald and P. M. Platzman, 
Phys. Rev. B33, 2481 (1986).



\bibitem{W}P. Wiegmann Phys. Rev. Lett. 108, 206810 (2012).


\bibitem{HV} For developments on this subject   see \cite{Avron,ReadHV,Tokatly,Son,Abanov}.






\bibitem{Landau}L.D. Landau, JETP11,542(1941);J.Phys. 5,71; 8,1(1941).
\bibitem{Feynman}R. P. Feynman,
 Phys. Rev. 91, 1291, 1301 (1953); 94, 262 (1954); R. P. Feynman, M. Cohen {\it ibid}. 102, 1189 (1956).

\bibitem{Kivelson}S. C. Zhang, T. H. Hansson,  S. A. Kivelson, Phys.
Rev. Lett. 62, 82 (1989).
\bibitem{Read89}
N. Read, Phys. Rev. Lett. 62, 86 (1989).

\bibitem{Dung}D-H. Lee, S. C. Zhang, Phys. Rev Lett 66, 1220 (1991).

\bibitem{Stone}M. Stone, Phys. Rev. B42, 212 (1990).


\bibitem{C}Our approach is a close analog of the hydrodynamics description  of Calogero model of Refs. \cite{Calogero}.
 \bibitem{Newton}V. V. Kozlov, General Theory of Vortices, Springer 2003.
\bibitem {Calogero}
 A.G.Abanov, P.Wiegmann, Phys.\ Rev. Lett.\  95:076402 (2005); E.Bettelheim, A.G.Abanov, P.Wiegmann, {\it ibid} 97:246401 (2006);\! M.Stone \!et\! al  J.Phys.A \!41:275401 (2008).

\bibitem{Bargmann}V. Bargmann, Rev. Mod. Phys. 34, 829 (1962).


\bibitem{GJ}S. M. Girvin, T. Jach, Phys. Rev. 8 29, 5617 (1984).






\bibitem{shift2}
X.-G. Wen, A. Zee, Phys. Rev. Lett. 69, 953 (1992). 

\bibitem{Avron}J. E. Avron, R. Seiler, P. G. Zograf, Phys. Rev. Lett.
75, 697 (1995).

\bibitem{Tokatly}I. V.
Tokatly, G. Vignale, Phys. Rev. B 76, 161305
(2007); J. Phys. C 21, 275603
(2009).

\bibitem{ReadHV}N. Read,
 Phys. Rev. B 79, 045308 (2009); 
 N. Read, E.H. Rezayi,
  Phys. Rev. B 84, 085316
(2011).

\bibitem{Son}C. Hoyos, D. T. Son, Phys. Rev. Lett. 108, 066805 (2012).
\bibitem{Abanov} A.G. Abanov, On the effective hydrodynamics of FQHE, to be published.






\bibitem{GZ}
Y. Greenberg, V. Zelevinsky, J.Phys.A 45, 035001 (2012).
 

\bibitem{electrostatics}  Coulomb forces always emerge
in  flows of   electronic liquids. They are essential factor  in the bulk,
less  essential on the edge. In this paper we neglect Coulomb 
forces in order to  unmask laws of  quantum hydrodynamics.  \bibitem{comment} In connection to FQHE the relation between density and
vorticity in a rotated Euler fluid was discussed in \cite{Stone}. Recently
the chiral compressible flow has been studied in \cite{Abanov,GZ}. Gapped
\(\propto q\Omega\)  linear
waves  have been observed.
\bibitem{Calogerocomment} A similar shift of velocity has been also observed  in hydrodynamics of Calogero model
\cite{Calogero}.
\bibitem{correltations}Eq. \eq{v3} expresses  flux  in terms of the density and the two-points density correlation function.
At the ground state where \(P=0\) the equation establishes a relation between the  density and two-point function obtained in  A. Zabrodin, P. Wiegmann, J.Phys. A39:8933 (2006).
\bibitem{comment1}   Incidently  a similar equation exists  inside the core. There the quantum corrections changes  the last term to \(-\frac{ 1}{4\pi} \nu\Delta\log\rho\).
Accidently   a similar equation  followed from  the
effective action of Refs.\cite {Kivelson,
Read89}  erroneously featuring the term  \(-\frac{ 1}{4\pi} \nu\Delta\log\rho\) inside and outside
of the vortex.
 \bibitem{Feinberg}Hydrodynamics of a non-chiral
version of the same model 
and with \(\beta=1\) has been studied in relation to normal matrices by J.
Feinberg, Nucl. Phys. B  705:403 (2005).
\bibitem{KirchhoffHamiltonian}The Hamiltonian proposed by Kirchhoff himself was \(\mathcal{H}=2\Omega\int\rho\Psi d^2z=-q\Omega\sum_i\log|z_i-z_j|^2\).
\bibitem {normalordering}In  formulas for the energy and momentum flux of Sec. 11 we assume the normal ordering of the density operator as in \cite{GMP} \(:\rho:=\sum_i\sum_ke^{-i\ell^2{\rm k}\p_{z_i}}e^{-i\frac 12\bar{\rm k}z_i}\). The ordering assures that the energy vanishes in the integer case \(\beta=1\).  
 \bibitem{Jack}We
would like to clarify the meaning of the Hamiltonian
 \eqref{V}.
The first quantized version of the Hamiltonian is \(H=\frac{1}{2m}\sum_{i=1}^Na^\dag_ia_i\)
is the second order differential operator, where \(a_i=2\hbar i(-\p_i+\beta\sum_{j\neq i}\frac{1}{z_i-z_j})=m\vv_i+\p_i\log|\psi_0|^2 \) \cite{W}. It acts in the space of symmetric
holomorphic polynomials multiplied by the factor \(\prod_{i>j}\!(z_i\!-\!\!z_j\!)^\beta\)
with the inner product \eq{B}. The Hamiltonian is a complexified version of the chiral sector  of Calogero model.  If \(Q_\lambda\)
is a basis in a ring of polynomials the Hamiltonian can be viewed as a  
 semi-infinite matrix   \(H_{\lambda'\lambda}\!=\!(2\hbar^2/m)\!\sum_i\!\int
\overline{\p_{z_i} Q_{\lambda^\prime}}\p_{z_i} Q _{\lambda}\prod_{i>j}\!|z_i\!-\!z_j|^{2\beta}
 d\mu\). We see that the Hamiltonian can be diagonalized in terms of  bi-orthogonal
polynomials with the weight \(\prod_{i>j}\!|z_i\!-\!z_j|^{2\beta}
 d\mu\).
These polynomials are bi-orthogonal analog of Jack polynomials. Contrary
to Jack polynomials, not much is known about their bi-orthogonal extension.




  




\bibitem{commentK} It is instructive
to analyze the   energy difference of the flow and the vortex flow. Simple computations yield \(
\\H_E-H=\frac{1}{2m}\int[-\frac{q}{2}  \nabla\times   P+\frac{q}{2}\rho\omega+(\frac
q 2  \nabla\sqrt\rho)^2] d^2 r 
\).\\   The  differential   \(\nabla\!\times\!
P\)  yields to the Lorentz shear force. 
 \bibitem{commentCS}  Lagrangian form of the hydrodynamics is not particularly helpful since subtle quantum phenomena and the chiral constraint are hidden in the measure of integration. However, it may
  help to compare our approach with
those of \cite{Kivelson, Read89,Abanov}. Let us  choose the density
of vortices \(\rho\) and its momentum \(\pi_\rho\) as canonical variables
and represent the velocity as \(m_*  v=  \nabla\pi_\rho-  a+\frac{q}{4} 
\nabla^*\log\rho\)  with a constraint \(-\nabla\times a=2\pi q(\rho-\bar\rho)\)
to match
\eq{vv}. The constraint  enforced by the Lagrangian multiplier \(a_0\) yields
the Chern-Simons term in the Lagrangian:  \(
L=-\rho(\dot\pi_\rho+ a_0)+a_0\bar\rho-\frac{1}{2\pi
q}a_0(\nabla\times
a-\frac{1}{2m_*}[\rho|\nabla\pi_\rho-  a|^2-q\rho( 
\nabla \times   a)+(
\frac {q}{2} \nabla\sqrt\rho)^2  
\)]. Two last terms differ this Lagrangian from those of \cite{Kivelson, Read89}.
They were recovered in \cite{Abanov}. The Lagrangian  can be expressed in terms of the vertex operator \(\Phi=\sqrt\rho e^{-\frac{\ii}{\hbar}\pi_\rho}\): \(L=a_0\bar\rho-\Phi^\ast[(i\hbar\p_t+a_0)\Phi-\frac{1}{2m_*}[|(i\hbar \nabla+  a)\Phi|^2+q( 
\nabla \times   a)|\Phi|^2+(\frac {q^2}{4}-1)(
 \nabla|\Phi|)^2]\). Incompressibility condition\(\) and the chiral constraint must be added.
 \bibitem{bi-hamiltonian}The transformation \eq{shift} is  analogues to the
Miura transformation of the theory of solitons.  
Similarly two Hamiltonian structures are analogues to  the bi-Hamiltonian
structure
of theory of solitons C. S. Gardner, J. Math. Phys. 12, 1548 (1971).  
 \bibitem{Virasoro} Elsewhere we discuss the relation of the extended algebra \eq{PP} and the Virasoro algebra of conformal field theory with the central charge \(c=1-6(\sqrt\beta-1/\sqrt\beta)^2\) (discussion with E. Bettelheim is acknowledged). 
\smallskip

\end{thebibliography}
\end{document}